\documentclass[10pt]{llncs}

\mainmatter
\usepackage{hyperref}
\usepackage{wrapfig} 
\usepackage{stmaryrd} %\llbracket for \sem
\usepackage{paralist}
\usepackage{algorithm}
\usepackage[noend]{algpseudocode}
\usepackage{listings}
\usepackage{cite}
\usepackage{todonotes}

\usepackage{color}
\usepackage{colortbl}

\input{macros}
\definecolor{StriverBlue}{RGB}{42, 0, 255}
\definecolor{StriverGreen}{RGB}{0, 155, 0}
\definecolor{StriverRed}{RGB}{155, 0, 0}
\definecolor{StriverOrange}{RGB}{255, 183, 68}
\definecolor{StriverYellow}{RGB}{155, 155, 45}
\lstdefinelanguage{Striver}
{
    basicstyle=\ttfamily,
    keywordstyle=[1]\color{StriverYellow},
    keywordstyle=[2]\color{StriverBlue},
    keywordstyle=[3]\bfseries,
    keywordstyle=[4]\color{StriverBlue},
    keywordstyle=[5]\color{StriverOrange},
    keywordstyle=[6]\color{StriverGreen},
    keywordstyle=[7]\color{StriverRed},
    otherkeywords = {
    <~,<<,~>,>>,macro,:=,const,input,output,true,false,ticks,define,outside,notick,U,delay,||,&&,+,-,==,!,!=,if,let,in
    then,else,int,bool,unit,Time,num,TV_Status},
    morekeywords = [1]{<~,<<,~>,>>},
    morekeywords = [2]{macro, :=, input,const, output, ticks, define},
    morekeywords = [3]{outside, notick, true, false},
    morekeywords = [4]{if, then, else},
    morekeywords = [5]{int, bool, unit, Time, void, num, TV_Status},
    morekeywords = [6]{U, delay, ||, &&, +, -, !, !=, ==},
    morekeywords = [7]{let, in},
}
\lstdefinelanguage{LOLA}
{
    escapechar={\~}, 
    basicstyle=\ttfamily,
    keywordstyle=[1]\color{StriverYellow},
    keywordstyle=[2]\color{StriverBlue},
    keywordstyle=[3]\bfseries,
    keywordstyle=[4]\color{StriverBlue},
    keywordstyle=[5]\color{StriverOrange},
    keywordstyle=[6]\color{StriverGreen},
    keywordstyle=[7]\color{StriverRed},
    otherkeywords = {macro,:=,const,input,output,true,false,define,outside,U,||,&&,+,-,==,!,!=,if,let,in then,else,int,bool,unit,Time,num,TV_Status},
    morekeywords = [1]{},
    morekeywords = [2]{macro, :=, input, const, output,  define},
    morekeywords = [3]{true, false},
    morekeywords = [4]{if, then, else},
    morekeywords = [5]{int, bool, unit, num, TV_Status},
    morekeywords = [6]{U, ||, &&, +, -, !, !=, ==, or, and, not, implies},
    morekeywords = [7]{let, in},
}

\usepackage{multirow}
\usepackage{wrapfig}

\graphicspath{ {figures/} } %so the figures can be included using just the filename and not the relative path

\usepackage{tabularx}
\usepackage{marvosym}
\usepackage[T1]{fontenc}

\def\orcidID#1{\smash{\href{http://orcid.org/#1}{\protect\raisebox{-1.25pt}{\protect\includegraphics{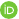}}}}}

\newcommand{\Thanks}{}

\title{Retroactive Parametrized Monitoring\Thanks}
\author{
Paloma Pedregal\inst{1,2}\textsuperscript{(\Letter)} \and
Felipe Gorostiaga\inst{1,3}\textsuperscript{(\Letter)}\orcidID{0000-0002-3478-3408} \and
C\'esar S\'anchez\inst{1}\orcidID{0000-0003-3927-4773}}
\institute{IMDEA Software Institute, Spain
  \and
  Universidad Polit\'ecnica de Madrid (UPM), Spain
  \and
  CIFASIS, Argentina
}

\begin{document}
\maketitle

\begin{abstract}
  In online monitoring, we first synthesize a monitor from a formal
  specification, which later runs in tandem with the system under
  study, incrementally receiving its progress and evolving with the
  system.
  In offline monitoring the trace is logged as the system progresses
  to later do post-mortem analysis after the system has finished
  executing.

  In this paper we propose \emph{retroactive dynamic
    parametrization}, a technique that allows a monitor to revisit
  the past log as it progresses, while still executing in an online
  manner.
  This feature allows new monitors to be incorporated into a running
  system and to revisit the past for particular behaviors based on new
  information discovered.
  Retroactive parametrization also allows a monitor to lazily ignore
  events and revisit and process them later, when it discovers that it
  should have followed those events.
  We showcase the use of retroactive dynamic parametrization to
  monitor denial of service attacks on a network using network logs.
  % 
  % We also show how to use this novel technique in the context of
  %blockchain, where the running system preserves the past, and where
  %monitors are created in the middle of the execution.
%
\end{abstract}

% Keywords:
% TODO

\section{Introduction}
%\felitext{Runtime Verification, SRV}
%

%% In this paper we introduce retroactive dynamic parametrization in
%% runtime verification.
%% %
%% We describe this technique in the context of stream runtime
%% verification, a functional implementation and its application to
%% network traffic analysis.

Runtime verification (RV) is a lightweight formal dynamic verification
technique that analyzes a single trace of execution using a monitor
derived from a specification
(see~\cite{leucker09brief,bartocci18lectures}).
The initial specification languages to describe monitors in RV where
borrowed from property languages for static verification, including
linear temporal logic (LTL)~ \cite{pnueli77temporal}, adapted to
finite traces
\cite{bauer06monitoring,manna92temporal}
or extensions~\cite{eisner03reasoning}.
Most RV languages describe a monolithic monitor that later processes the
events received.
Dynamic parametrization (also known as parametric trace slicing) allows
quantifying over objects and spawn monitors that follow
independently the actual objects observed, like in Quantified Event
Automata (QEA)~\cite{barringer12quantified}.

Stream runtime verification
\cite{goodloe10monitoring,pike10copilot,bozzelli16foundations}(SRV),
pioneered by \lola \cite{dangelo05lola}
defines monitors by declaring the dependencies between output streams
and input streams.
The initial application domain of \lola{} was the testing of
synchronous hardware.
Temporal testers~\cite{pnueli06psl} were later proposed as a
monitoring technique for LTL based on Boolean streams.
Copilot~\cite{pike10copilot,goodloe10monitoring,perez20copilot} is a
DSL that, similar to \lola, declares dependencies between streams in a
Haskell-based style, to generate C monitors.
Lola2.0~\cite{faymonville16stream} extends \lola allowing
dynamically parametrized streams, similarly to QEA.
Stream runtime verification has also been extended recently to
asynchronous and real-time
systems~\cite{faymonville17realtime,convent18tessla,leucker18tessla,gorostiaga18striver}.
HLola\cite{ceresa20declarative,gorostiaga21hlola,gorostiaga21stream}
is an implementation of \lola as an embedded DSL in Haskell, which
allows borrowing datatypes from Haskell directly as \lola expressions, and
using features like higher-order functions to ease the development of
specifications and the runtime system.
In this paper we use HLola and extend it with capabilities for
retroactive dynamic parametrization.

In some situations we want to monitor a property that we define once
the system is already running, and we cannot or do not wish to stop
and restart the monitor.
%
% One example is monitors that observe the evolution of a blockchain
% that has been running for a long time before the monitor was written
% and compiled.
%
Therefore, the monitor will receive online the events of a suffix of
the original trace.
At this point, we must decide to either
(1) \emph{ignore} that the monitor is plugged in the middle of the
trace and let it work under the premise that it is watching the whole
history,
(2) \emph{encode the lack of knowledge} about the beginning of the
trace in the specification, or,
(3) if the beginning of the trace was preserved in a \emph{log} or a
database and it is accessible, allow the monitor to collaborate with
the log to process the missing past and then continue to process the
future events online.
The first option is the most natural and in many cases a good
solution, while the second option has been explored in
\cite{gorostiaga22monitorability}, but these two options ignore the
beginning of the trace and thus we have to pick one of them if the
trace is not preserved.
The third option proposes a novel combination of \emph{offline+online
  monitoring}, which offers the possibility of accessing the past
of the trace.
Moreover, enriching an SRV monitor with the ability of accessing the
past allows the description of properties that revisit the past
exploiting information discovered at a later time.

In this paper we show an instance of this combination of online and
offline monitoring to implement a novel dynamic instantiation of
monitors called \emph{retroactive dynamic parametrization} that
combines dynamic parametrization with the ability to revisit the past
of a live stream of events.

This paper is structured as follows.
Section~\ref{sec:prelims} introduces SRV and static parametrization,
Section~\ref{sec:offnline} presents the syntax and semantics  of the
operators that enable retroactive dynamic parametrization,
Section~\ref{sec:implementation} shows how we have implemented these
features in our tool, Section~\ref{sec:empirical} includes an
empirical evaluation of the tool, and finally
Section~\ref{sec:conclusion} concludes.

% \paragraph{Related work.}
% \felitext{
% Mainly, \lolatwo implements dynamic parametrization, but not
% retroactive.
% %
% In DBMSs the queries are stateless and there are no guarantees wrt
% resource requirements.
% %
% Online and offline RV in general?
% }

%%% Local Variables:
%%% TeX-master: "../main.tex"
%%% TeX-PDF-mode: t
%%% End:

\section{Preliminaries}
\label{sec:prelims}
Stream Runtime Verification (SRV) generalizes monitoring algorithms to
arbitrary data, where temporal dependencies are maintained but
datatypes are generalized using multi-sorted interpreted signatures,
which we call data theories.
The signatures are interpreted in the sense that every functional
symbol $\texttt{f}$ used to construct terms of a given type is
accompanied with an evaluation function $f$ (the interpretation) that
allows the computation of values (given values of the arguments).
In this paper we use the extensible tool \hlola, an implementation of
\lola~\cite{dangelo05lola} developed as an embedded DSL in Haskell.
We show in Section~\ref{sec:implementation} how \hlola allows the easy
implementation of new powerful features as libraries with no changes to
the core engine, leveraging the extensibility and expressiveness of
the data theories of the tool.
In this paper we show how retroactive parametrization is implemented
in \hlola.

A \lola specification $\tupleof{I,O,E}$ consists of
\begin{inparaenum}[(1)]
\item a set of typed input stream variables $I$, which correspond to
  the inputs observed by the monitor;
\item a set of typed output stream variables $O$ which represent the
  outputs of the monitor as well as intermediate observations; and
\item defining equations, which associate every output $y\in O$ with a
  stream expression $E_y$ that describes declaratively the intended
  values of $y$ (in terms of the values of inputs and output streams).
\end{inparaenum}
The set of \emph{stream expressions} of a given type is built from
constants and function symbols as constructors (as usual), and also
from \emph{offset expressions} of the form $s$\LOLAACC{$k$}{$d$} where $s$ is a
stream variable, $k$ is an integer number and $d$ is a value of the
type of $s$ (used as default).
For example, \LOLACODE{altitude\LOLAACC{-1}{0.0}} represents the value
of stream \LOLACODE{altitude} in the previous step of time, with
\LOLACODE{0.0} as the value used at the initial instant.
We can define a stream \STRNAME{alt\_ok} which checks that the
\LOLACODE{altitude} is always below a predefined threshold, for example
\LOLACODE{100} meters:

\noindent {\frontendsize\tt
\begin{tabularx}{\textwidth}{@{}r@{\hspace{.5em}}l}
\multicolumn{2}{X}{\color[HTML]{888888}\rule[2mm]{30mm}{.1pt}} \\
\linensize{\color[HTML]{4d798f}1} & \textbf{input} \textcolor[HTML]{008000}{\textbf{Double}} \textcolor[HTML]{eb7600}{\textbf{altitude}} \\
\linensize{\color[HTML]{4d798f}2} & \textbf{output} \textcolor[HTML]{008000}{\textbf{Bool}} \textcolor[HTML]{eb7600}{\textbf{alt\_ok}} \textbf{=} altitude\textbf{[now]} < 100\\
\multicolumn{2}{X}{\color[HTML]{888888}\rule[2mm]{30mm}{.1pt}}
\end{tabularx}}

Given values of the input streams, the formal semantics of a \lola
specification is defined denotationally as the unique collection of
streams of values that satisfies all equations.
The existence and uniqueness of streams of outputs are guaranteed when
specifications are not cyclic.

Moreover, online efficient monitoring algorithms can be synthesized
for specifications with (bounded) future
accesses~\cite{dangelo05lola,sanchez18online}.
Efficiency here means that resources (time and space) are independent
of the length of the trace and can be calculated statically, and that
each individual output can be generated immediately.
% \felitodo{Caution: this is modulo data theories. We later show graphs
% whose space is length-dependent and cannot be precalculated}
%
\hlola can be efficiently monitored in this trace-length independent
sense~\cite{ceresa20declarative}.
As a byproduct of its design, \hlola allows \emph{static
  parametrization} in stream definitions, this is, streams that
abstract away some concrete values, which can be instantiated by the
compiler.
Following the previous example, we can define a more generic version
of the stream \STRNAME{alt\_ok} as follows: 

\noindent {\frontendsize\tt
\begin{tabularx}{\textwidth}{@{}r@{\hspace{.5em}}l}
\multicolumn{2}{X}{\color[HTML]{888888}\rule[2mm]{30mm}{.1pt}} \\
\linensize{\color[HTML]{4d798f}1} & \textbf{input} \textcolor[HTML]{008000}{\textbf{Double}} \textcolor[HTML]{eb7600}{\textbf{altitude}} \\
\linensize{\color[HTML]{4d798f}2} & \textbf{output} \textcolor[HTML]{008000}{\textbf{Bool}} \textcolor[HTML]{eb7600}{\textbf{alt\_checker}} \textbf{<}\textcolor[HTML]{008000}{\textbf{Double}} \textcolor[HTML]{96325d}{\textbf{threshold}}\textbf{>} \textbf{=} altitude\textbf{[now]} < threshold \\
\linensize{\color[HTML]{4d798f}3} & \textbf{output} \textcolor[HTML]{008000}{\textbf{Bool}} \textcolor[HTML]{eb7600}{\textbf{alt\_ok}} \textbf{=} (altchecker 100)\textbf{[now]}\\
\multicolumn{2}{X}{\color[HTML]{888888}\rule[2mm]{30mm}{.1pt}}
\end{tabularx}}

% \label{ex:paramaltitude} Removed because we cannot refer to it

Static parametrization does not allow instantiations of a parametric
stream with a value that is discovered at runtime.
In static parametrization parameters must be known in static time
before the monitor starts running.
In the following sections we will detail the reason of this limitation
and we offer alternatives to implement dynamic parametrization,
analysing their pros and cons.
\subsubsection{Nested Monitors.}
The keystone of the design of \hlola is to use datatypes and functions
from Haskell as the data theories of \lola.
In turn, \hlola also allows using \lola specifications as datatypes,
via a function \reserved{runSpec} that executes a specification
over the input trace and produces a value of the result type.
As a consequence, we can use \lola as a data theory within \lola
itself, an approach that we have called \emph{nested
  monitors}~\cite{gorostiaga21nested}.
Nested monitors allows writing functions on streams as SRV
specifications, creating and executing these specifications
dynamically.
Nested specifications are particularly useful when the caller monitor
can invoke a nested monitor passing a sub-trace of the original trace
as input.
The \hlola operator $s$\LOLASLICE{$n$} creates a list with the next
$n$ values of the stream $s$.
We often use slices as input streams for nested specifications, but
any list of values of the appropriate type can be used.
The type of the stream \STRNAME{$x$} determines the type of the value
returned when the specification is invoked dynamically.

Defining a \emph{nested} specification involves giving it a name and
adding an extra clause: \RETURNWHEN{$x$}{$y$} where \STRNAME{$x$} is a
stream of any type and \STRNAME{$y$} is a \TYNAME{Boolean} stream.
The type of the stream \STRNAME{$x$} determines the type of the value
returned when the specification is invoked dynamically.
Once we have defined a nested specification, we can execute it using
the function \LOLACODE{runSpec}.

\begin{qedexample}
  The following specification calculates whether input numeric streams
  \STRNAME{r} and \STRNAME{s} will cross within the following $50$
  instants.
We define a topmost specification as follows, where stream
\STRNAME{willCross} invokes the nested specification
\INNERNAME{crossspec} with the slices containing the next $50$ events
of \STRNAME{r} and \STRNAME{s} as input.

\noindent {\frontendsize\tt
\begin{tabularx}{\textwidth}{@{}r@{\hspace{.5em}}l}
\multicolumn{2}{X}{\color[HTML]{888888}\rule[2mm]{30mm}{.1pt}} \\
\linensize{\color[HTML]{4d798f}1} & \textbf{use innerspec }\textcolor[HTML]{4d8491}{\textbf{crossspec}} \\[0.5em]
\linensize{\color[HTML]{4d798f}2} & \textbf{input} \textcolor[HTML]{008000}{\textbf{Double}} \textcolor[HTML]{eb7600}{\textbf{r}} \\
\linensize{\color[HTML]{4d798f}3} & \textbf{input} \textcolor[HTML]{008000}{\textbf{Double}} \textcolor[HTML]{eb7600}{\textbf{s}} \\
\linensize{\color[HTML]{4d798f}4} & \textbf{output} \textcolor[HTML]{008000}{\textbf{Bool}} \textcolor[HTML]{eb7600}{\textbf{willCross}} \textbf{=} runSpec (crossspec r\textbf{[:}50\textbf{]} s\textbf{[:}50\textbf{]})\\
\multicolumn{2}{X}{\color[HTML]{888888}\rule[2mm]{30mm}{.1pt}}
\end{tabularx}}

\noindent In our example, the nested specification \INNERNAME{crossspec} is:

\noindent {\frontendsize\tt
\begin{tabularx}{\textwidth}{@{}r@{\hspace{.5em}}l}
\multicolumn{2}{X}{\color[HTML]{888888}\rule[2mm]{30mm}{.1pt}} \\
\linensize{\color[HTML]{4d798f}1} & \textbf{innerspec }\textcolor[HTML]{008000}{\textbf{Bool}} \textcolor[HTML]{4d8491}{\textbf{crossspec}}  \\[0.5em]
\linensize{\color[HTML]{4d798f}2} & \textbf{input} \textcolor[HTML]{008000}{\textbf{Double}} \textcolor[HTML]{eb7600}{\textbf{r}} \\
\linensize{\color[HTML]{4d798f}3} & \textbf{input} \textcolor[HTML]{008000}{\textbf{Double}} \textcolor[HTML]{eb7600}{\textbf{s}} \\
\linensize{\color[HTML]{4d798f}4} & \textbf{output} \textcolor[HTML]{008000}{\textbf{Bool}} \textcolor[HTML]{eb7600}{\textbf{cross}} \textbf{=} \\
\linensize{\color[HTML]{4d798f}5} & \phantom{..}sign (r\textbf{[now]} - s\textbf{[now]}) /= sign (r\textbf{[}-1\textbf{|}r\textbf{[now]}\textbf{]} - s\textbf{[}-1\textbf{|}s\textbf{[now]}\textbf{]}) \\[0.5em]
\linensize{\color[HTML]{4d798f}6} & \textcolor[HTML]{4d8491}{\textbf{return }}\textcolor[HTML]{eb7600}{\textbf{cross}}\textcolor[HTML]{4d8491}{\textbf{ when }}\textcolor[HTML]{eb7600}{\textbf{cross}}\\
\multicolumn{2}{X}{\color[HTML]{888888}\rule[2mm]{30mm}{.1pt}}
\end{tabularx}}

\noindent{}The output stream \STRNAME{cross} simply checks that the
relative order of the streams \STRNAME{r} and \STRNAME{s} changes.
A nested specification with a clause \RETURNWHEN{$x$}{$y$} returns the
value of \STRNAME{$x$} at the first time \STRNAME{$y$} becomes \TRUE,
or the last value of \STRNAME{$x$} if \STRNAME{$y$} never holds.
Therefore, if \STRNAME{$y$} becomes \TRUE in the middle of an
execution, the monitor does not have to run until the end to compute a
value and can anticipate the result, so nested specifications can be
incrementally computed as new elements of the input slice are
available, and return the outcome as soon as it is definite.
\end{qedexample}
%

%%% Local Variables:
%%% TeX-master: "../main.tex"
%%% TeX-PDF-mode: t
%%% End:

\section{Combinig Online and Offline Monitoring}
\label{sec:offnline}
We introduce features to combine offline and online runtime
verification.
We start with nested monitors that can access past events when new
information is discovered.

\subsection{Retroactive Nested Monitors}
%
%\begin{changed}
% Original text:We plan to use nested monitors to inspect the past upon request.
%\end{changed}
%
\begin{qedexample}
  Consider monitoring network traffic, where the monitor receives the
  source and destination of an IP packet, and the packets per second
  in the last hundred instants.
  We want to detect whether an address has received too many packets
  in the last hundred instants, specified as follows:
  \textit{if the packet flow is low, then there is no attack, but when
  the flow rate is above a predefined threshold
  (\LOLACODE{threshold\_pps}) we have to inspect the last hundred
  packets and check if a given address is under attack.}
We use the following specification to check the heuristics and
trigger the finer analysis of the tail of the trace when necessary.

\noindent {\frontendsize\tt
\begin{tabularx}{\textwidth}{@{}r@{\hspace{.5em}}l}
\multicolumn{2}{X}{\color[HTML]{888888}\rule[2mm]{30mm}{.1pt}} \\
\linensize{\color[HTML]{4d798f}1} & \textbf{input} \textcolor[HTML]{008000}{\textbf{Int}} \textcolor[HTML]{eb7600}{\textbf{packets\_per\_second}} \\[0.5em]
\linensize{\color[HTML]{4d798f}2} & \textbf{output} \textcolor[HTML]{008000}{\textbf{Int}} \textcolor[HTML]{eb7600}{\textbf{counter}} \textbf{=} counter\textbf{[}-1\textbf{|}0\textbf{]} + 1 \\
\linensize{\color[HTML]{4d798f}3} & \textbf{output} \textcolor[HTML]{008000}{\textbf{Bool}} \textcolor[HTML]{eb7600}{\textbf{under\_attack}} \textbf{=} \\
\linensize{\color[HTML]{4d798f}4} & \phantom{..}\textcolor[HTML]{1c21b8}{if} packets\_per\_second\textbf{[now]} > threshold\_pps \textcolor[HTML]{1c21b8}{then} \\
\linensize{\color[HTML]{4d798f}5} & \phantom{..}\phantom{..}\textcolor[HTML]{1c21b8}{let} past = pastRetriever counter\textbf{[now]} \textcolor[HTML]{1c21b8}{in} \\
\linensize{\color[HTML]{4d798f}6} & \phantom{..}\phantom{..}runSpec (finerSpec `withTrace` past) \\
\linensize{\color[HTML]{4d798f}7} & \phantom{..}\textcolor[HTML]{1c21b8}{else} False\\
\multicolumn{2}{X}{\color[HTML]{888888}\rule[2mm]{30mm}{.1pt}}
\end{tabularx}}

\noindent The auxiliary stream \STRNAME{counter} indicates
\LOLACODE{pastRetriever} which events it has to recover.
We use \IFTHENELSEPH instead of the Boolean operator
\LOLACODE{(\&\&)} to stress the fact that the nested
specification is only executed when the threshold is exceeded.
Then, we define the nested specification \INNERNAME{finerSpec} as
follows:

\noindent {\frontendsize\tt
\begin{tabularx}{\textwidth}{@{}r@{\hspace{.5em}}l}
\multicolumn{2}{X}{\color[HTML]{888888}\rule[2mm]{30mm}{.1pt}} \\
\linensize{\color[HTML]{4d798f}1} & \textbf{innerspec }\textcolor[HTML]{008000}{\textbf{Bool}} \textcolor[HTML]{4d8491}{\textbf{finerSpec}}  \\
\linensize{\color[HTML]{4d798f}2} & \textbf{input} \textcolor[HTML]{008000}{\textbf{String}} \textcolor[HTML]{eb7600}{\textbf{destination}} \\[0.5em]
\linensize{\color[HTML]{4d798f}3} & \textbf{output} \textcolor[HTML]{008000}{\textbf{(Map String Int)}} \textcolor[HTML]{eb7600}{\textbf{entropy}} \textbf{=} \\
\linensize{\color[HTML]{4d798f}4} & \phantom{..}insertWith (+) destination\textbf{[now]} 1 entropy\textbf{[}-1\textbf{|}empty\textbf{]} \\[0.5em]
\linensize{\color[HTML]{4d798f}5} & \textbf{output} \textcolor[HTML]{008000}{\textbf{Bool}} \textcolor[HTML]{eb7600}{\textbf{under\_attack}} \textbf{=} maximum (elems entropy\textbf{[now]}) > threshold\_dest \\[0.5em]
\linensize{\color[HTML]{4d798f}6} & \textcolor[HTML]{4d8491}{\textbf{return }}\textcolor[HTML]{eb7600}{\textbf{under\_attack}}\textcolor[HTML]{4d8491}{\textbf{ when }}\textcolor[HTML]{eb7600}{\textbf{under\_attack}}\\
\multicolumn{2}{X}{\color[HTML]{888888}\rule[2mm]{30mm}{.1pt}}
\end{tabularx}}

\noindent The function \LOLACODE{insertWith cmb k v m} inserts the
element \LOLACODE{v} associated to the key \LOLACODE{k} into the map
\LOLACODE{m}, combining \LOLACODE{v} with the previous value
\LOLACODE{m[k]} using \LOLACODE{cmb} if it exists.
The function \LOLACODE{elems m} returns the list of values in the map
\LOLACODE{m}, and the function \LOLACODE{maximum ls} returns the
maximum in the list \LOLACODE{ls}.
The nested specification will return \TRUE as soon as
\STRNAME{under\_attack} becomes \TRUE, and \FALSE if such stream is
always \FALSE until the end of the (sub)trace.
\end{qedexample}
Note how this specification detects an attack at most one hundred
instants after it happens.
Also, the nested monitors in this example are created, executed and
destroyed at every instant.
In Section~\ref{sec:implementation}, we explain how we can implement
dynamic parametrization by keeping the nested monitors alive across
the trace.
\subsection{(Forward) Dynamic Parametrization}
We introduce now a novel operator \OVER, which lets us instantiate a
parametric stream over dynamic values.
The \OVER operator takes a parametric stream \STRNAME{strm} of type
\TYNAME{S} with a parameter of type \TYNAME{P}, and a stream
\STRNAME{params} of sets of values of type \TYNAME{P}, and creates
an expression of type \TYNAME{Map P S}, whose keys at any given
instant are the values in \LOLACODE{params\NOW}, and where
%
% \felitodo{unless it
%  has never passed the filter of \LOLACODE{updating}
%  Cesar: yes, but ``technical detail''.
%}
%
the value associated to each key is the instantiation of
\STRNAME{s} over the key.

Using the \OVER operator we can dynamically instantiate a parametric
stream over a set of parameters that are discovered while processing
the trace of input.
\begin{qedexample}
\label{ex:3wayhandshake}
The specification below checks that every pair \PARAMNAME{(source,
  destination)} follows the TCP $3$-way handshake in which (1) the
source sends a packet \LOLACODE{SYN}, then (2) the destination sends
\LOLACODE{SYN/ACK} and then (3) the source sends \LOLACODE{ACK}.
The following state machine described in \hlola captures this:

\noindent {\frontendsize\tt
\begin{tabularx}{\textwidth}{@{}r@{\hspace{.5em}}l}
\multicolumn{2}{X}{\color[HTML]{888888}\rule[2mm]{30mm}{.1pt}} \\
\linensize{\color[HTML]{4d798f}1} & \textbf{input} \textcolor[HTML]{008000}{\textbf{String}} \textcolor[HTML]{eb7600}{\textbf{source}} \\
\linensize{\color[HTML]{4d798f}2} & \textbf{input} \textcolor[HTML]{008000}{\textbf{String}} \textcolor[HTML]{eb7600}{\textbf{destination}} \\
\linensize{\color[HTML]{4d798f}3} & \textbf{input} \textcolor[HTML]{008000}{\textbf{PacketType}} \textcolor[HTML]{eb7600}{\textbf{p\_type}} \\[0.5em]
\linensize{\color[HTML]{4d798f}4} & \textbf{data }\textcolor[HTML]{008000}{\textbf{PacketType }}\textbf{=} SYN | SYNACK | ACK \\
\linensize{\color[HTML]{4d798f}5} & \phantom{..}\phantom{..}\phantom{..}\phantom{..}\phantom{..}\phantom{..}\phantom{..} \textbf{deriving} (Generic,Read,FromJSON,Eq) \\
\linensize{\color[HTML]{4d798f}6} & \textbf{data }\textcolor[HTML]{008000}{\textbf{State }}\textbf{=} Uninit | Error | SYNED | SYNACKED \\
\linensize{\color[HTML]{4d798f}7} & \phantom{..}\phantom{..}\phantom{..}\phantom{..}\phantom{..}\phantom{..}\phantom{..} \textbf{deriving} (Generic,Read,FromJSON,Eq) \\[0.5em]
\linensize{\color[HTML]{4d798f}8} & \textbf{output} \textcolor[HTML]{008000}{\textbf{State}} \textcolor[HTML]{eb7600}{\textbf{state}} \textbf{<}\textcolor[HTML]{008000}{\textbf{(String, String)}} \textcolor[HTML]{96325d}{\textbf{srcdest}}\textbf{>} \textbf{=} \textcolor[HTML]{1c21b8}{let} \\
\linensize{\color[HTML]{4d798f}9} & \phantom{..}prev\_state = (state srcdest)\textbf{[}-1\textbf{|}Uninit\textbf{]} \\
\linensize{\color[HTML]{4d798f}10} & \phantom{..}\textcolor[HTML]{1c21b8}{in} \\
\linensize{\color[HTML]{4d798f}11} & \phantom{..}\textcolor[HTML]{1c21b8}{if} srcdest /= (source\textbf{[now]}, destination\textbf{[now]}) \textcolor[HTML]{1c21b8}{then} prev\_state \\
\linensize{\color[HTML]{4d798f}12} & \phantom{..}\textcolor[HTML]{1c21b8}{else} \textcolor[HTML]{1c21b8}{if} p\_type\textbf{[now]} == SYN\phantom{..}\phantom{..}\&\& prev\_state == Uninit\phantom{..} \textcolor[HTML]{1c21b8}{then} SYNED \\
\linensize{\color[HTML]{4d798f}13} & \phantom{..}\textcolor[HTML]{1c21b8}{else} \textcolor[HTML]{1c21b8}{if} p\_type\textbf{[now]} == SYNACK \&\& prev\_state == SYNED\phantom{..}\phantom{..}\textcolor[HTML]{1c21b8}{then} SYNACKED \\
\linensize{\color[HTML]{4d798f}14} & \phantom{..}\textcolor[HTML]{1c21b8}{else} \textcolor[HTML]{1c21b8}{if} p\_type\textbf{[now]} == ACK\phantom{..}\phantom{..}\&\& prev\_state == SYNACKED \textcolor[HTML]{1c21b8}{then} Uninit \\
\linensize{\color[HTML]{4d798f}15} & \phantom{..}\textcolor[HTML]{1c21b8}{else} Error\\
\multicolumn{2}{X}{\color[HTML]{888888}\rule[2mm]{30mm}{.1pt}}
\end{tabularx}}

\noindent
However, we cannot know statically for which pair of addresses we have
to instantiate this parametrized stream.
Therefore, we define a set of parameters (in this case, a set of pairs
of addresses) that we need to follow:

\noindent {\frontendsize\tt
\begin{tabularx}{\textwidth}{@{}r@{\hspace{.5em}}l}
\multicolumn{2}{X}{\color[HTML]{888888}\rule[2mm]{30mm}{.1pt}} \\
\linensize{\color[HTML]{4d798f}16} & \textbf{output} \textcolor[HTML]{008000}{\textbf{(Set (String, String))}} \textcolor[HTML]{eb7600}{\textbf{params}} \textbf{=} \\
\linensize{\color[HTML]{4d798f}17} & \phantom{..}insert (source\textbf{[now]}, destination\textbf{[now]}) params[-1,empty]\\
\multicolumn{2}{X}{\color[HTML]{888888}\rule[2mm]{30mm}{.1pt}}
\end{tabularx}}

\noindent
At every instant, we add the current pair \LOLACODE{(source\NOW,
destination\NOW)} to the set of parameters.
We are now ready to parametrize the parametric stream \STRNAME{state}
over the values of \STRNAME{params}:

\noindent {\frontendsize\tt
\begin{tabularx}{\textwidth}{@{}r@{\hspace{.5em}}l}
\multicolumn{2}{X}{\color[HTML]{888888}\rule[2mm]{30mm}{.1pt}} \\
\linensize{\color[HTML]{4d798f}18} & \textbf{output} \textcolor[HTML]{008000}{\textbf{Bool}} \textcolor[HTML]{eb7600}{\textbf{all\_ok}} \textbf{=} \textcolor[HTML]{1c21b8}{let} \\
\linensize{\color[HTML]{4d798f}19} & \phantom{..}states = elems (state `over` params) \\
\linensize{\color[HTML]{4d798f}20} & \phantom{..}\textcolor[HTML]{1c21b8}{in} all (/= Error) states\\
\multicolumn{2}{X}{\color[HTML]{888888}\rule[2mm]{30mm}{.1pt}}
\end{tabularx}}

\noindent
Note that we simply check that all the values in the map generated by
the \OVER expression are different from \LOLACODE{Error}, this is,
that no pair of addresses is in the \LOLACODE{Error} state.
This specification follows each pair of addresses independently.
One thing to note is that we never remove parameters from
\STRNAME{params}, and as a consequence, the set can grow indefinitely,
even though it is not necessary.
In fact, we can remove the previous pair of source and destination
when the previous type of message was \LOLACODE{ACK}.
The reason why we keep the parametric stream alive for one more
instant before removing it is to let \STRNAME{all\_ok} check that the
behavior was correct.
\end{qedexample}
Every time a new value is incorporated to the set of parameters, we
spawn a new monitor parametrized with the new value.
Then, we preserve the state of this monitor between instants in an
auxiliary stream, executing the nested monitor alongside the outer
monitor until the auxiliary monitor is no longer needed, that is,
until its associated parameter is removed from the set.

The nested monitors just described will process the same events as the
root monitor, but it is often the case that only some of the events
are relevant to a specific parametrized stream.
We can use \emph{subtracing} to only update a subset of the
parametrized streams at every instant.
\subsection{Subtracing}
Parametrized streams are internal monitors that observe the same
trace as their parent.
However, it is often the case that parametrized streams only care
about the events that are related to their parameter, as in
Example~\ref{ex:3wayhandshake}, where the events irrelevant to an
instantiation do not change its state.
Subtracing lets the nested stream focus only on its relevant subtrace.
Moreover, subtracing enables the possibility of updating only the
nested monitors that may be affected at the current instant.
To implement subtracing we introduce the operator \UPDATING that lets
us specify the parameters of the monitors that have to process the
current event.

For example, we can use subtracing in Example~\ref{ex:3wayhandshake}
redefining the streams \STRNAME{state} and \STRNAME{all\_ok} as
follows:

\noindent {\frontendsize\tt
\begin{tabularx}{\textwidth}{@{}r@{\hspace{.5em}}l}
\multicolumn{2}{X}{\color[HTML]{888888}\rule[2mm]{30mm}{.1pt}} \\
\linensize{\color[HTML]{4d798f}8} & \textbf{output} \textcolor[HTML]{008000}{\textbf{State}} \textcolor[HTML]{eb7600}{\textbf{state}} \textbf{<}\textcolor[HTML]{008000}{\textbf{(String, String)}} \textcolor[HTML]{96325d}{\textbf{srcdest}}\textbf{>} \textbf{=} \textcolor[HTML]{1c21b8}{let} \\
\linensize{\color[HTML]{4d798f}9} & \phantom{..}prev\_state = (state srcdest)\textbf{[}-1\textbf{|}Uninit\textbf{]} \\
\linensize{\color[HTML]{4d798f}10} & \phantom{..}\textcolor[HTML]{1c21b8}{in}\phantom{..} \textcolor[HTML]{1c21b8}{if} p\_type\textbf{[now]} == SYN\phantom{..}\phantom{..}\&\& prev\_state == Uninit\phantom{..} \textcolor[HTML]{1c21b8}{then} SYNED \\
\linensize{\color[HTML]{4d798f}11} & \phantom{..}\textcolor[HTML]{1c21b8}{else} \textcolor[HTML]{1c21b8}{if} p\_type\textbf{[now]} == SYNACK \&\& prev\_state == SYNED\phantom{..}\phantom{..}\textcolor[HTML]{1c21b8}{then} SYNACKED \\
\linensize{\color[HTML]{4d798f}12} & \phantom{..}\textcolor[HTML]{1c21b8}{else} \textcolor[HTML]{1c21b8}{if} p\_type\textbf{[now]} == ACK\phantom{..}\phantom{..}\&\& prev\_state == SYNACKED \textcolor[HTML]{1c21b8}{then} Uninit \\
\linensize{\color[HTML]{4d798f}13} & \phantom{..}\textcolor[HTML]{1c21b8}{else} Error \\[0.5em]
\linensize{\color[HTML]{4d798f}18} & \textbf{output} \textcolor[HTML]{008000}{\textbf{Bool}} \textcolor[HTML]{eb7600}{\textbf{all\_ok}} \textbf{=} \textcolor[HTML]{1c21b8}{let} \\
\linensize{\color[HTML]{4d798f}19} & \phantom{..}currentpair = singleton (source\textbf{[now]}, destination\textbf{[now]}) \\
\linensize{\color[HTML]{4d798f}20} & \phantom{..}states = elems (state `over` params `updating` currentpair) \\
\linensize{\color[HTML]{4d798f}21} & \phantom{..}\textcolor[HTML]{1c21b8}{in} all (/= Error) states\\
\multicolumn{2}{X}{\color[HTML]{888888}\rule[2mm]{30mm}{.1pt}}
\end{tabularx}}

\begin{figure}[t]
 \centering
 \includegraphics{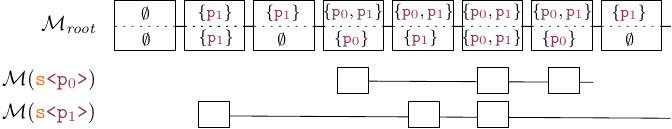}
 \caption{Subtracing of a parametrized stream \STRNAME{s} over
 parameters \PARAMNAME{p$_0$} and \PARAMNAME{p$_1$}. At every instant
 of the \reserved{root} monitor we show the
 sets of dynamic parameters on top, and the sets of parameters to
 update below.}
 \label{fig:subtracing}
\end{figure}

\noindent
Using subtracing, we can separate the parent trace from the traces
received by the nested monitors, who will only observe a subtrace of
the original trace.
This is depicted in Fig.~\ref{fig:subtracing}.

In the example above, the monitor follows the dynamically parametrized
stream once the parameter has been discovered.
In fact, the monitor may not even have access to the past of the trace
because the online runtime monitoring infrastructure erases past
events once they are processed.
The only alternative is to monitor a parametrized stream after its
parameter is discovered (like in
\lolatwo~\cite{faymonville19streamlab}).

If the infrastructure keeps the past events of the trace, we can
combine \emph{retroactive nested monitors} with \emph{dynamic
  parametrization} when a new parameter is discovered, effectively
implementing \emph{retroactive dynamic parametrization}.

\subsection{Retroactive Dynamic Parametrization}
Static parametrization in \hlola is limited because the monitor
cannot know the state of a parametric stream over an arbitrary
parameter in the middle of the trace unless the parameter is
determined statically before the monitor starts running.
One alternative to circumvent this limitation is to statically
instantiate the stream with every potentially interesting parameter.
This strategy is reasonable if the stream can only be parametrized
with a domain of parameters with few elements, like Boolean or a small
enumerated type, but it becomes unfeasible when the space of potential
parameters is large.

Instead, we propose an alternative technique: that the monitor
revisits the past of the trace every time it discovers a new parameter
to instantiate the parametric stream and from that point continue to
monitor the new stream online.

To implement this feature we add a special clause \WITHINIT to the
\OVER operator that allows specifying an \emph{initializer}, which
indicates how the nested monitor can update its state up to the current
instant.
An initializer will typically call an external program with the
corresponding arguments that indicate an offline infrastructure how to
efficiently retrieve the elements of past of the trace that are
relevant to the parameter.
%
% At the time of initialization, the initializer can use this
% infrastructure to access the values of other streams at past points in
% time.
% \felitodo{I don't get the last sentence}

\begin{figure}[t]
 \centering
 \includegraphics{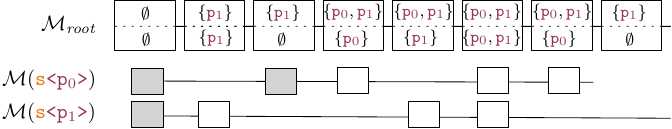}
 \caption{Subtracing and retroactive effects in a parametrized stream
   \STRNAME{s} over \PARAMNAME{p$_0$} and \PARAMNAME{p$_1$}. We show
   the sets of dynamic parameters on top, a the parameters to update
   below. The events in grey are those retrieved by the corresponding
   initializer.}
 \label{fig:retrovsfwdtraces}
\end{figure}

\begin{qedexample}
\label{ex:openfiles}
Suppose we are monitoring file accesses in a file system, and we want
to assess that every time a file is read or written, it must have been
created previously.
We could use \emph{forward dynamic parametrization} as in
Example~\ref{ex:3wayhandshake} to follow all open files.

\noindent {\frontendsize\tt
\begin{tabularx}{\textwidth}{@{}r@{\hspace{.5em}}l}
\multicolumn{2}{X}{\color[HTML]{888888}\rule[2mm]{30mm}{.1pt}} \\
\linensize{\color[HTML]{4d798f}1} & \textbf{input} \textcolor[HTML]{008000}{\textbf{String}} \textcolor[HTML]{eb7600}{\textbf{file\_id}} \\
\linensize{\color[HTML]{4d798f}2} & \textbf{input} \textcolor[HTML]{008000}{\textbf{OpType}} \textcolor[HTML]{eb7600}{\textbf{operation}} \\[0.5em]
\linensize{\color[HTML]{4d798f}3} & \textbf{data }\textcolor[HTML]{008000}{\textbf{State }}\textbf{=}\phantom{..}Created | NE | Error \textbf{deriving} (Generic,Read,FromJSON,Eq) \\
\linensize{\color[HTML]{4d798f}4} & \textbf{data }\textcolor[HTML]{008000}{\textbf{OpType }}\textbf{=} Create\phantom{..}| RW \textbf{deriving} (Generic,Read,FromJSON,Eq) \\[0.5em]
\linensize{\color[HTML]{4d798f}5} & \textbf{output} \textcolor[HTML]{008000}{\textbf{State}} \textcolor[HTML]{eb7600}{\textbf{state}} \textbf{<}\textcolor[HTML]{008000}{\textbf{String}} \textcolor[HTML]{96325d}{\textbf{fid}}\textbf{>} \textbf{=} \textcolor[HTML]{1c21b8}{let} \\
\linensize{\color[HTML]{4d798f}6} & \phantom{..}prev\_state = (state fid)\textbf{[}-1\textbf{|}NE\textbf{]} \textcolor[HTML]{1c21b8}{in} \\
\linensize{\color[HTML]{4d798f}7} & \phantom{..}\textcolor[HTML]{1c21b8}{if} (operation\textbf{[now]} == Create \&\& prev\_state == NE) \\
\linensize{\color[HTML]{4d798f}8} & \phantom{..}|| (operation\textbf{[now]} == RW\phantom{..}\phantom{..} \&\& prev\_state == Created) \\
\linensize{\color[HTML]{4d798f}9} & \phantom{..}\textcolor[HTML]{1c21b8}{then} Created \textcolor[HTML]{1c21b8}{else} Error \\[0.5em]
\linensize{\color[HTML]{4d798f}10} & \textbf{output} \textcolor[HTML]{008000}{\textbf{(Set String)}} \textcolor[HTML]{eb7600}{\textbf{params}} \textbf{=} insert file\_id\textbf{[now]} params[-1,empty] \\[0.5em]
\linensize{\color[HTML]{4d798f}11} & \textbf{output} \textcolor[HTML]{008000}{\textbf{Bool}} \textcolor[HTML]{eb7600}{\textbf{all\_ok}} \textbf{=} \textcolor[HTML]{1c21b8}{let} \\
\linensize{\color[HTML]{4d798f}12} & \phantom{..}fid = singleton file\_id\textbf{[now]} \\
\linensize{\color[HTML]{4d798f}13} & \phantom{..}states = elems (state `over` params `updating` fid) \\
\linensize{\color[HTML]{4d798f}14} & \phantom{..}\textcolor[HTML]{1c21b8}{in} all (/= Error) states\\
\multicolumn{2}{X}{\color[HTML]{888888}\rule[2mm]{30mm}{.1pt}}
\end{tabularx}}

\noindent
In this example we define datatypes to capture the state of a file id
(\LOLACODE{Created} or \LOLACODE{NE}, ``non existent'') and also the
type of operations (\LOLACODE{Create} or \LOLACODE{RW}, ``read or
write'').
Just like in Example~\ref{ex:3wayhandshake}, we define the state
machine and we check that all the values are different from
\LOLACODE{Error}.
\end{qedexample}
Although this specification is valid, if most of the created files are
never read or written (or only some read or writes are relevant for
the specification), this forward monitor would be following many open
files unnecessarily.
% \felitodo{Considering malicious users is interesting, but not part of
% this example, as it is}
%
\begin{qedexample}
  
  We can use \emph{retroactive} monitoring to only create the dynamic
  parameters when the corresponding file id is read or written, and
  use the retroactive capability to inspect the past of the trace and
  see if the file was created previously.
  We redefine the streams \STRNAME{params} and \STRNAME{all\_ok}
  accordingly.

\noindent {\frontendsize\tt
\begin{tabularx}{\textwidth}{@{}r@{\hspace{.5em}}l}
\multicolumn{2}{X}{\color[HTML]{888888}\rule[2mm]{30mm}{.1pt}} \\
\linensize{\color[HTML]{4d798f}10} & \textbf{output} \textcolor[HTML]{008000}{\textbf{(Set String)}} \textcolor[HTML]{eb7600}{\textbf{params}} \textbf{=} \textcolor[HTML]{1c21b8}{let} \\
\linensize{\color[HTML]{4d798f}11} & \phantom{..}prev\_params = params[-1,empty] \\
\linensize{\color[HTML]{4d798f}12} & \phantom{..}\textcolor[HTML]{1c21b8}{in} \textcolor[HTML]{1c21b8}{if} operation == RW \textcolor[HTML]{1c21b8}{then} \\
\linensize{\color[HTML]{4d798f}13} & \phantom{..}\phantom{..}insert file\_id\textbf{[now]} prev\_params \\
\linensize{\color[HTML]{4d798f}14} & \phantom{..}\textcolor[HTML]{1c21b8}{else} prev\_params \\[0.5em]
\linensize{\color[HTML]{4d798f}15} & \textbf{output} \textcolor[HTML]{008000}{\textbf{Bool}} \textcolor[HTML]{eb7600}{\textbf{all\_ok}} \textbf{=} \textcolor[HTML]{1c21b8}{let} \\
\linensize{\color[HTML]{4d798f}16} & \phantom{..}fid = singleton file\_id\textbf{[now]} \\
\linensize{\color[HTML]{4d798f}17} & \phantom{..}states = elems (state `over` params `updating` fid `withInit` initer) \\
\linensize{\color[HTML]{4d798f}18} & \phantom{..}\textcolor[HTML]{1c21b8}{in} all (/= Error) states\\
\multicolumn{2}{X}{\color[HTML]{888888}\rule[2mm]{30mm}{.1pt}}
\end{tabularx}}

\noindent The new stream of parameters only adds a file id when it is
read or written.
The new \OVER expression specifies an initializer \LOLACODE{initer}
(whose definition is not shown in the specification) that calls an
external program to retrieve the past of the corresponding parameter.
The external program may use an index to efficiently retrieve only the
events in the past relevant to the current file id.
In fact, the program could fetch only the \LOLACODE{open} events
associated with the file id, making the installation phase even
faster.
\end{qedexample}
%
% \felitext{Here I was starting to talk about initializers, saying that
% they were useful to outsource the computation of the state of a
% monitor, and perhaps that it could be used to save the state of the
% root monitor and load it later, but I don't think it is too
% interesting for this paper}
%
Non-retroactive dynamic parametrization avoids retrieving old events
at the expense of not preserving the semantics of static
parametrization.
Instead, the parametrized stream considers that the trace only
started when the parameter was first instantiated shows up.
We depict this conceptual difference in
Fig.~\ref{fig:retrovsfwdtraces}.
\paragraph{Maybe over.} We now describe additional operators that
introduce useful syntactic sugar.
The operator \MOVER (called \emph{maybe over}) allows the discovery of
a single parameter in the middle of the trace.
Instead of applying a parametrized stream over a stream of sets of
values, the operator \MOVER takes the parametric stream and a stream
of type \TYNAME{Maybe P}, creating an expression of type \TYNAME{Maybe
S}.
As in the original \OVER, if it is retroactive or not depends on the
presence of \WITHINIT.
This syntactic sugar simply creates a stream of sets of parameters
which will contain at most one element at every instant, and returns
the only value in the map when the map is not empty.

\begin{qedexample}
\label{ex:dynaltitude}
To illustrate \MOVER, we instantiate the specification in
Section~\ref{sec:prelims} with a parametric threshold that is received
in the middle of the trace.

\noindent {\frontendsize\tt
\begin{tabularx}{\textwidth}{@{}r@{\hspace{.5em}}l}
\multicolumn{2}{X}{\color[HTML]{888888}\rule[2mm]{30mm}{.1pt}} \\
\linensize{\color[HTML]{4d798f}1} & \textbf{input} \textcolor[HTML]{008000}{\textbf{Double}} \textcolor[HTML]{eb7600}{\textbf{altitude}} \\
\linensize{\color[HTML]{4d798f}2} & \textbf{input} \textcolor[HTML]{008000}{\textbf{Double}} \textcolor[HTML]{eb7600}{\textbf{threshold}} \\[0.5em]
\linensize{\color[HTML]{4d798f}3} & \textbf{output} \textcolor[HTML]{008000}{\textbf{Bool}} \textcolor[HTML]{eb7600}{\textbf{alt\_checker}} \textbf{<}\textcolor[HTML]{008000}{\textbf{Double}} \textcolor[HTML]{96325d}{\textbf{threshold}}\textbf{>} \textbf{=} altitude\textbf{[now]} < threshold \\[0.5em]
\linensize{\color[HTML]{4d798f}4} & \textbf{output} \textcolor[HTML]{008000}{\textbf{(Maybe Double)}} \textcolor[HTML]{eb7600}{\textbf{mthreshold}} \textbf{=} \textcolor[HTML]{1c21b8}{let} \\
\linensize{\color[HTML]{4d798f}5} & \phantom{..}prev\_mthreshold = mthreshold\textbf{[}-1\textbf{|}Nothing\textbf{]} \\
\linensize{\color[HTML]{4d798f}6} & \phantom{..}\textcolor[HTML]{1c21b8}{in} \textcolor[HTML]{1c21b8}{if} prev\_mthreshold == Nothing \&\& threshold\textbf{[now]} > 0 \\
\linensize{\color[HTML]{4d798f}7} & \phantom{..}\phantom{..} \textcolor[HTML]{1c21b8}{then} Just threshold\textbf{[now]} \\
\linensize{\color[HTML]{4d798f}8} & \phantom{..}\phantom{..} \textcolor[HTML]{1c21b8}{else} prev\_mthreshold \\[0.5em]
\linensize{\color[HTML]{4d798f}9} & \textbf{output} \textcolor[HTML]{008000}{\textbf{Bool}} \textcolor[HTML]{eb7600}{\textbf{alt\_ok}} \textbf{=} \textcolor[HTML]{1c21b8}{let} \\
\linensize{\color[HTML]{4d798f}10} & \phantom{..}mok = alt\_checker `mover` mthreshold \\
\linensize{\color[HTML]{4d798f}11} & \phantom{..}\textcolor[HTML]{1c21b8}{in} mok /= Just False\\
\multicolumn{2}{X}{\color[HTML]{888888}\rule[2mm]{30mm}{.1pt}}
\end{tabularx}}

\noindent
The value of \STRNAME{mthreshold} will be the first positive value of
\STRNAME{threshold}, and \LOLACODE{Nothing} until then.
The stream \STRNAME{alt\_ok} will be \FALSE only when the threshold
has been set and the current altitude is greater than the established
threshold.
We can use the operator \WITHINIT to add an initializer and also
revisit the past of the trace to see if the altitude had been exceeded
before the actual value was determined.
\end{qedexample}
% Note that when the stream \STRNAME{threshold} is
% \LOLACODE{th}$\neq$\LOLACODE{Nothing}, the value of the expression
% \LOLACODE{altitude `over` threshold} is exactly the same as the value
% of the statically parametrized stream \LOLACODE{altitude th}.
%
\paragraph{Subtrace filter with expression.} The operator \WHEN lets
us replace the \UPDATING clause with an expression that indicates when
a parameter should be updated.
For example, we can specify which file id to update using the \WHEN
operator in Example~\ref{ex:openfiles} as follows:

\noindent {\frontendsize\tt
\begin{tabularx}{\textwidth}{@{}r@{\hspace{.5em}}l}
\multicolumn{2}{X}{\color[HTML]{888888}\rule[2mm]{30mm}{.1pt}} \\
\linensize{\color[HTML]{4d798f}18} & \textbf{output} \textcolor[HTML]{008000}{\textbf{Bool}} \textcolor[HTML]{eb7600}{\textbf{all\_ok}} \textbf{=} \textcolor[HTML]{1c21b8}{let} \\
\linensize{\color[HTML]{4d798f}19} & \phantom{..}states = elems (state `over` params `when` (== file\_id\textbf{[now]})) \\
\linensize{\color[HTML]{4d798f}20} & \phantom{..}\textcolor[HTML]{1c21b8}{in} all (/= Error) states\\
\multicolumn{2}{X}{\color[HTML]{888888}\rule[2mm]{30mm}{.1pt}}
\end{tabularx}}

\noindent The syntactic sugar will calculate the set of parameters
$\{\LOLACODE{p} \text{ such that } e(\LOLACODE{p})\}$, where $e$ is
the expression associated to the \WHEN clause.
Note that if we use \WHEN, the engine has to check the condition $e$
over every live parameter to see if the corresponding stream has to be
updated.
It is more efficient to directly specify the set of parameters to
update, if possible.

%%% Local Variables:
%%% TeX-master: "../main.tex"
%%% TeX-PDF-mode: t
%%% End:

\section{Implementation}
\newcommand{\PARAMP}{\PARAMNAME{$p$}}
\label{sec:implementation}
The implementation of the novel operators in \hlola is based on the
rich type expressivity of the tool, which include the ability to
handle the \lola language itself as a data theory within
\hlola~\cite{gorostiaga21nested}.

Dynamically mapping a parametric stream \STRNAME{strm} with a stream
of set of parameters \STRNAME{params} of type \TYNAME{Set P}, creates
an auxiliary stream \STRNAME{x\_over\_params} of type \TYNAME{Map P
FrozenMonitor} that associates, at every instant, each parameter
\PARAMP in
\STRNAME{params} with the state of the nested monitor corresponding to
\LOLACODE{s\PARAMNAME{<\PARAMP{}>}}.
The value of \LOLACODE{x `over` params} is simply the projection of
the parametrized streams in the frozen monitors of
\LOLACODE{x\_over\_params\NOW}.
There are three possibilities for the behavior of the auxiliary
stream for a given parameter \PARAMP:
\begin{compactenum}[(1)]
\item $\PARAMP \in
  \LOLACODE{params\LOLAACC{-1}{$\emptyset$}} \setminus
  \LOLACODE{params\NOW}$: the parameter was in the set in the previous
  instant, but it is no longer in the set in the current instant.
In this case, \PARAMP and its associated value are deleted
from the map \LOLACODE{x\_over\_params\LOLAACC{-1}{$\emptyset$}}.
\item $\PARAMP \in
  \LOLACODE{params\LOLAACC{-1}{$\emptyset$}} \cap
  \LOLACODE{params\NOW}$: the parameter was in the set and it is still
  in the set now.
%
%  We first check whether \PARAMP belongs to the set of parameters
%  indicated by \LOLACODE{updating}.
%
  If \PARAMP is in \LOLACODE{updating}, we feed the current event to the
  monitor associated with \PARAMP and let it progress one step.
  Then, the value of the parametrized stream in the nested monitor is
  associated with \PARAMP in the returned map.
\item
  $\PARAMP \in \LOLACODE{params\NOW} \setminus
  \LOLACODE{params\LOLAACC{-1}{$\emptyset$}}$: the parameter was not
  in the set, but it is now.
  The monitor for \PARAMP is installed, executing the initializer
  (possibly revisiting the past) to get the monitor up to date and
  ready to continue online.
  Once the monitor is installed, we proceed as in the previous case,
  checking whether \PARAMP is in \LOLACODE{updating} to decide to
  inject the current event, and associating the value of
  \LOLACODE{s<\PARAMP{}>} to \PARAMP in the returned map.
  Note that the past is only revisited when a new parameter is
  discovered.
  Once the stream is instantiated with the parameter, its
  corresponding nested monitor will continue executing over the future
  of the trace online.
\end{compactenum}

A nested monitor can be installed with an empty initializer and at the
same time \PARAMP may not be in \LOLACODE{updating}.
In this case the nested monitor has not processed any event, the value
of \LOLACODE{s<\PARAMP{}>} is non existent and the sets
\LOLACODE{params\NOW} and
$\reserved{keys}(\LOLACODE{x\_over\_params\NOW})$ differ.

Since we want \hlola to support initialization from different sources
(e.g. a DBMS, a blockchain node, or plain log files) the
initializer of the internal monitors typically invokes an external
program.
This external program, called \emph{adapter}, is in charge of
recovering the trace and formatting it adequately for the monitor.
%
% Fig.~\ref{fig:architecture} shows the architecture of our approach.
The following is the architecture of our approach.

%\begin{figure}[b!]
  \centerline{\includegraphics[scale=0.97]{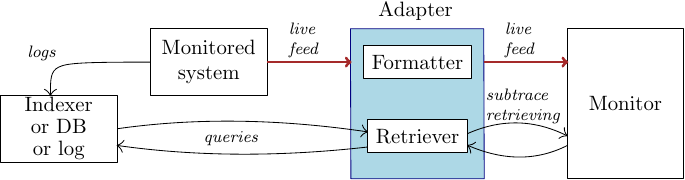}}
%  \caption{Architecture of \hlola with retroactive capabilities}
%\label{fig:architecture}
%\end{figure}

%
% We also propose to use the adapter to retrieve the trace of the
% program online, represented by the bold red lines with label ``Live feed''
%in the figure.
%
% \begin{wrapfigure}[6]{r}{0.50\textwidth} 
%   % \vspace{-4mm}
%   % \hspace{-1.5em}
%   % \begin{tabular}{c}
%   \includegraphics{src/figures/vertarchitecture.pdf}
% \end{tabular}
% \caption{Architecture of \hlola with retroactive capabilities}
% \label{fig:architecture}
% \end{wrapfigure}
%

%%% Local Variables:
%%% TeX-master: "../main.tex"
%%% TeX-PDF-mode: t
%%% End:

\newcommand{\KW}[1]{\textbf{(#1)}\xspace}
\newcommand{\Hone}{\KW{H1}}
\newcommand{\Htwo}{\KW{H2}}
\newcommand{\Hthree}{\KW{H3}}
\newcommand{\Hfour}{\KW{H4}}
\newcommand{\Hfive}{\KW{H5}}
\newcommand{\Hsix}{\KW{H6}}
\newcommand{\Done}{\KW{D1}}
\newcommand{\Dtwo}{\KW{D2}}
\newcommand{\Dthree}{\KW{D3}}
\newcommand{\Dfour}{\KW{D4}}
\newcommand{\Sone}{\KW{S1}}
\newcommand{\Stwo}{\KW{S2}}
\newcommand{\Sthree}{\KW{S3}}

\section{Empirical Evaluation}
\label{sec:empirical}
%\palotext{Revisar los tiempos de los verbos en toda la sección (use/used)}
We report in this section an empirical evaluation of retroactive
dynamic parame\-trization.
We have used HLola to implement several algorithms for the detection of
distributed denial of service attacks (DDOS).
All the experiments were executed on a Linux machine with $256$GB of
RAM and $72$ virtual cores (Xeon Gold 6154 @3GHz) running Ubuntu
20.04.
The monitors were generated using the implementation of
HLola\footnote{Available at
  \url{https://github.com/imdea-software/hlola/}} described
in~\cite{gorostiaga21hlola}.
For conciseness we use RP to refer to retroactive parametrization,
non-RP to implementations that do not use retroactive parametrization
(and use dynamic parametrization instead).
We evaluate empirically the following hypotheses:
\begin{itemize}
\item[\Hone] RP is functionally
  equivalent to non-RP.
\item[\Htwo] RP and non-RP run at similar speeds, particularly when
  most dynamic instantiations turn out to be irrelevant.
\item[\Hthree] RP consumes significantly less memory than non-RP,
  particularly when most instantiations are irrelevant.
\item[\Hfour] Aggregated RP---where the
  monitor receives summaries of the trace that indicate if further
  processing is necessary---is functionally equivalent to RP.
%\item[\Hfive] Checking a reported violation forward is cheaper
%  than finding a violation.
\item[\Hfive] Aggregated RP is much
  more efficient than RP and non-RP without aggregation.
\end{itemize}

The datasets for the experiments are (anonymized) samples of real
network traffic collected by a Juniper MX408 router that routes the
traffic of an academic network used by several tens of thousand of
users (students and researchers) simultaneously, routing approximately
$15$ Gbps of traffic on average.
The sampling ratio provided by the routers was $1$ to $100$
flows\footnote{Most detection systems use a much slower sampling of
  $1$ to $1000$ or even less.}
Each flow contains the metadata of the traffic sampled, with
information such as source and destination ports and addresses,
protocols and timestamps, but does not carry information about the
contents of the packets.
These flows are stored in aggregated batches of $5$ minutes encoded in
the \textbf{netflow} format.

Our monitors implement fourteen known DDOS network attacks detection
algorithms.
An attack is detected if the volume of connections to a destination
address surpasses a fixed attack-specific threshold, and those
connections come from a sufficiently large number of different
attackers, identified by source IP address.
The number of different source addresses communicating with a
destination is known as the \emph{entropy} of the destination.
Each attack is concerned with a different port and protocol and
considers a different entropy as potentially dangerous.

In order to process the network data needed by the monitors, we
developed a Python adapter that uses \texttt{nfdump}, a toolset to
collect and process netflow data.
The tool \texttt{nfdump} can be used to obtain all the flows in a
batch, optionally applying some simple filters, or to obtain
summarized information about all the flows in the batch.
For example, \texttt{nfdump} can provide all the flows received,
filtered by a protocol or address, as well as the volume of traffic to
the IP address that has received the most connections of a specific
kind.

In our empirical evaluation we use four datasets in which we knew
whether each attack was present:
\begin{itemize}
\item[\Done] One batch of network flows with an attack based on
  malformed UDP packets (UDP packets with destination port $0$).
  This batch contains $419938$ flows, with less than $1\%$ malformed
  UDP packets.
  The threshold for this attack is $2000$ packets per second, which is
  surpassed in this batch for one single address, for which the
  entropy of $5$ is exceeded.
\item[\Dtwo] One batch of network flows with no attack, containing
  $361867$ flows, of which only $66$ are malformed UDP packets
  (roughly, $0.001\%$).
\item[\Dthree] One batch of network flows with no attack, but with
  many origin IP addresses and $100$ destination addresses.  Based on
  D2, we modified the source and destination IP addresses.
\item[\Dfour] Intervals with several batches, where only one batch has
  an attack based on malformed UDP packets.
\end{itemize}

The monitors in our experiments follow the same attack description:
\emph{In a batch of 5 minutes of flow records, an address is under
  attack if it receives more than $t_0$ packets per second or bits per
  second from more than $t_1$ different source addresses
(where $t_0$ and $t_1$  depend on the attack).}
We have implemented monitors in three different ways\footnote{Due to
  space constraints, we have included the specifications for \Sone,
  \Stwo and \Sthree in Appendix~\ref{appendix}.  }
\begin{itemize}
\item[\Sone] Brute force: using dynamic parametrization the monitor
  calculates the number of packets and bits per second (which we call
  ``volume'') for all potential target IP addresses.
  It also computes the entropy for each potential target address and
  for each attack.
  For every flow, the monitor internally updates the information about
  the source address, destination and volume.

\item[\Stwo] Retroactive: the monitor also analyses all flows,
  calculating the volume of packets for each address, but in this case
  the monitor lazily avoids calculating the entropy.
  The monitor only calculates the entropy, using RP, when the volume
  of traffic for an address surpasses the threshold.
  The monitor uses the \emph{over} operator to revisit the past flows
  of the batch filtered by that attack, using the Python adapter which
  produces the subset of the flows required to compute the entropy.
  The monitoring then continues calculating the entropy until the end
  of the batch.
\item[\Sthree] Aggregated: the monitor receives a summary of a $5$
  minute batch of network data, as a single event containing $14$
  attack markers.
  The monitor is based on the ability of the backend to pre-process
  batches using \texttt{nfdump} to obtain---for each attack and for
  the whole batch---the maximum volume of traffic for any IP address.
  %
  % The monitor receives one single marker per attack after every batch.
  % 
  If an attack marker is over the threshold, the monitor spawns a
  nested monitor that retrieves a subsets of the flows for that batch
  and attack, which behaves as in \textbf{S2}.

  The aggregation of data provides a first, coarse overview serving as
  a necessary condition to spawn the expensive nested monitor.
  This is particularly useful because attacks are infrequent and the
  ratio of false positives of the summary detection is relatively low.
  %
  % If the address with most traffic surpasses the threshold, a 
  % in-depth process is necessary: it is possible that more than one
  % addresses are under attack, or that the address with most traffic
  % does not surpass the IP entropy threshold, but a different address
  % does surpass both.
  % %
  % If the volume threshold is exceeded for at least one IP address but
  % the IP entropy threshold is not exceeded by any IP, we say that the
  % early detection was a false positive.
\end{itemize}

In the first experiment we run the three implementations against
dataset \Dfour.
In this interval of multiple batches, only one of which contains an
attack, all three implementations identify the batch with the attack
and correctly detect the kind of attack and target address.
This confirms empirically hypotheses \Hone and \Hfour.

In the second experiment we run specifications \Sone and \Stwo against
datasets \Done, \Dtwo and \Dthree.
%
% The results are reported in Fig.~\ref{fig:exps}.
The results are reported in the following table:

% \begin{figure}[t!]
\begin{center}
  \begin{tabular}{ |c|c|c|c| } 
    \hline
    & \Done (Attack) & \Dtwo (No Attack) & \Dthree (No Attack)\\
    \hline
    \Sone (Brute force) & 18m12.146s & 15m51.599s & 16m34.795s \\
    \hline
    \Stwo (Retroactive) & 20m43.921s & 17m19.844s & 19m30.518s \\
    \hline
    \Sthree (Aggregated) & 0m16.208s & 0m2.109s & 0m2.115s \\ 
    \hline
  \end{tabular}
\end{center}
% \caption{Exp 2: Average running times of the implementations with D1, D2 and D3}
%  \label{fig:exps}
%\end{figure}   

\noindent We can see that the running times for the brute force and
retroactive implementations are similar, while the aggregated
implementation is extremely fast in comparison, which empirically
confirms \Htwo and \Hfive.
This is because \Sthree exploits the summarized information, and does
not find any marker over the threshold for the datasets \Dtwo and
\Dthree so the flows within the batch are never individually
processed.
For dataset \Done, a nested monitor will be executed because one of
the markers (for the attack with malformed UPD packets) is over the
threshold, but it will only try to detect the attack corresponding to
that marker, and it will only receive a small subset of the flows
(less than $1\%$ of the flows of the batch are relevant for the
attack).
If all the markers for all the attacks were over their threshold and
all the flows were implicated in the attacks, the time required would
be closer to the retroactive implementation.
The ad-hoc aggregation of data by the external tool is very efficient,
as is the verification of this data by the monitor, so this
implementation is especially advantageous when the positives (or false
positives) are expected to be infrequent, and when most of the data
can be filtered out before executing the nested monitor.

\begin{figure}[t!]
\centering
%% \begin{tabular}{@{}c@{\hspace{0.3em}}c@{\hspace{0.3em}}c@{}}
\begin{tabular}{@{}c@{}c@{}c@{}}
  \includegraphics[scale=0.6]{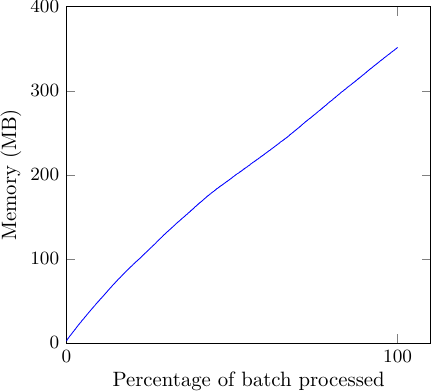} & \includegraphics[scale=0.6]{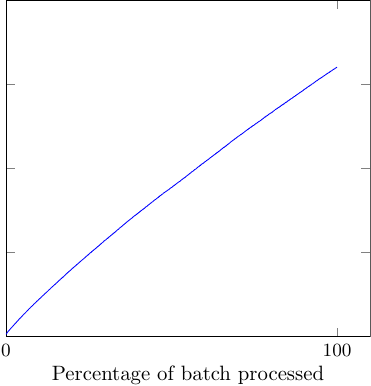} & \includegraphics[scale=0.6]{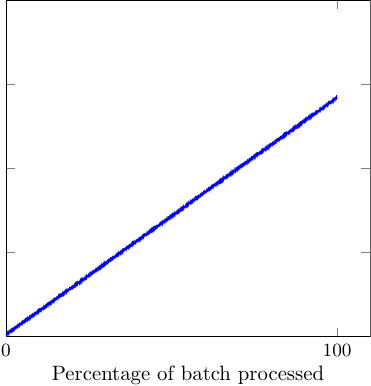} \\
  (a) \Sone with \Done & (b) \Sone with \Dtwo & (c) \Sone with \Dthree \\[1.2em]
  \includegraphics[scale=0.6]{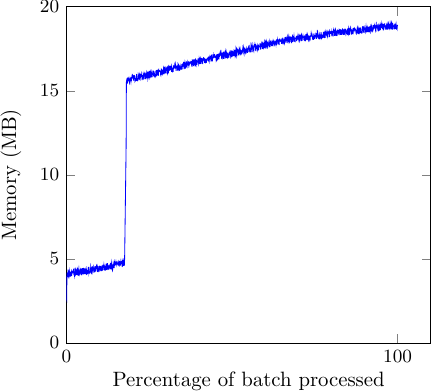} & \includegraphics[scale=0.6]{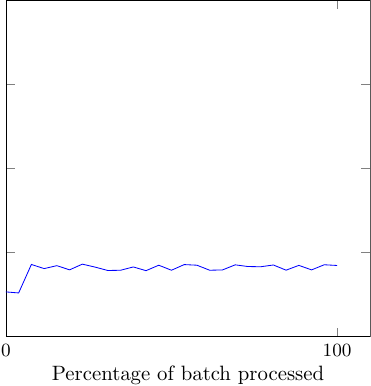} & \includegraphics[scale=0.6]{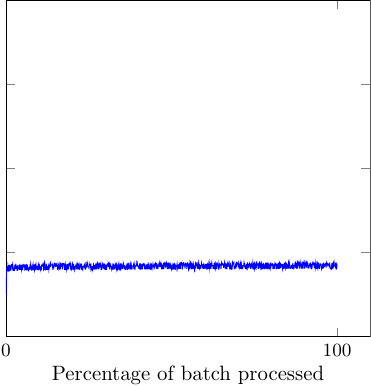} \\
  (d) \Stwo with \Done & (e) \Stwo with \Done & (f) \Stwo with \Dthree \\[1.2em]
%%   (d) Retroactive (S2)\\attack (D1) & (e) Retroactive (S2)\\no attack (D2) & (f) Retroactive (S2)\\no attack modified (D3) \\[1.2em]
\end{tabular}  
  \caption{Memory usage of the brute force (a), (b), (c) and retroactive (d), (e), (f).}
\label{fig:empiric:fut:plot} 
\end{figure}

In a third experiment we run a version of specifications \Sone and
\Stwo---instrumented with capabilities to measure memory
consumption---on \Done, \Dtwo and \Dthree.
The results, reported in Fig.~\ref{fig:empiric:fut:plot}, empirically
confirm \Hthree.
For the three datasets, the memory used by the brute force approach
increases linearly over time, as it has to keep track of the
volume and IP entropy for every attack and every potential target
address.
On the other hand, the memory usage of the retroactive implementation
remains close to constant, with a sudden increase when an attack is
detected and the past flows have to be retrieved and processed.

%% In a fourth experiment we run the instrumented version of the
%% aggregated implementation \Sthree.
%% %
%% The memory usage for dataset \Done, when there is an attack is
%% $216$MB, but for the datasets \Dtwo and \Dthree, when there is no
%% attack, the memory usage is $0$.
%% %
%% This result validates \Hfive in terms of memory usage.

%% \palotext{Advantages of Aggregation: most times the markers will not
%%   be over the threshold and the aggregation of data is very fast, thus
%%   avoiding unnecessary calculations. Make it clear that we could not
%%   ONLY use the preprocessing -> the ip with most traffic could not be
%%   the one with enough entropy, but if at least one address is over the
%%   volume threshold, any that are over the volume threshold will be
%%   analyzed}

%%% Local Variables:
%%% TeX-master: "../main.tex"
%%% TeX-PDF-mode: t
%%% End:

\section{Conclusions}
\label{sec:conclusion}
In this paper we have introduced the concept of retroactive dynamic
parametrization.
In dynamic parametrization, proposed in QEA and \lolatwo a new monitor
(which is an instance of a generic monitor) is instantiated the first time a
parameter is discovered.
In retroactive dynamic parametrization the decision to instantiate a
dynamic parametric monitor can be taken later in the future, for
example when a given parameter is discovered to be interesting.

Effectively implementing retroactive parametrization requires the
ability to revisit the history of the computation, a task that can be
efficiently implemented with a logging system.
Therefore, retroactive parametrization allows a fruitful combination
of offline and online monitoring.
Retroactive parametrization also allows monitors to be created in the
middle of an execution without requiring to process the whole trace
from the beginning.

We have implemented this technique in \hlola and empirically evaluated
its efficiency, illustrating that it can efficiently detect distributed
denial of service attacks in realistic network traffic.

%%% Local Variables:
%%% TeX-master: "../main.tex"
%%% TeX-PDF-mode: t
%%% End:

\vfill
\pagebreak

%\addcontentsline{toc}{chapter}{Bibliography}
%\bibliographystyle{apalike}
%\bibliographystyle{plain}
\bibliographystyle{splncs04}
\bibliography{biblio,papers}

\vfill
\pagebreak
\appendix
\section{Specifications of the Empirical Evaluation}
\label{appendix}
In this appendix we show the specifications that we have used to
empirically evaluate the different techniques of retroactive dynamic
parameterization in the context of network attack detection.

The first specification analyzes summaries of batches of flows,
executing a nested specification over the current batch (indicated by
\STRNAME{file\_id} and the suspected attack when one of the markers is
above a predefined threshold:

\noindent {\frontendsize\tt
\begin{tabularx}{\textwidth}{@{}r@{\hspace{.5em}}l}
\multicolumn{2}{X}{\color[HTML]{888888}\rule[2mm]{30mm}{.1pt}} \\
\linensize{\color[HTML]{4d798f}1} & \textbf{use innerspec }\textcolor[HTML]{4d8491}{\textbf{flowAnalyzer}} \\
\linensize{\color[HTML]{4d798f}2} & \textbf{input} \textcolor[HTML]{008000}{\textbf{String}} \textcolor[HTML]{eb7600}{\textbf{fileid}} \\
\linensize{\color[HTML]{4d798f}3} & \textbf{input} \textcolor[HTML]{008000}{\textbf{Int}} \textcolor[HTML]{eb7600}{\textbf{marker<AttackData>}} \\[0.5em]
\linensize{\color[HTML]{4d798f}4} & \textbf{output} \textcolor[HTML]{008000}{\textbf{[String]}} \textcolor[HTML]{eb7600}{\textbf{attacked\_IPs}} \textbf{=} map detect attacks \\
\linensize{\color[HTML]{4d798f}5} & \phantom{..}\textcolor[HTML]{1c21b8}{where} detect atk = (attack\_detection atk)\textbf{[now]} \\[0.5em]
\linensize{\color[HTML]{4d798f}6} & \textbf{define} \textcolor[HTML]{008000}{\textbf{String}} \textcolor[HTML]{eb7600}{\textbf{attack\_detection}} \textbf{<}\textcolor[HTML]{008000}{\textbf{AttackData}} \textcolor[HTML]{96325d}{\textbf{atk}}\textbf{>} \textbf{=}  \\
\linensize{\color[HTML]{4d798f}7} & \phantom{..}\textcolor[HTML]{1c21b8}{if} (marker atk)\textbf{[now]} > threshold atk \textcolor[HTML]{1c21b8}{then} \\
\linensize{\color[HTML]{4d798f}8} & \phantom{..}\phantom{..}runSpec (flowAnalyzer atk `withTrace` pastRetriever atk fileid\textbf{[now]}) \\
\linensize{\color[HTML]{4d798f}9} & \phantom{..}\textcolor[HTML]{1c21b8}{else} "No attack"\\
\multicolumn{2}{X}{\color[HTML]{888888}\rule[2mm]{30mm}{.1pt}}
\end{tabularx}}

\noindent{} The constant \LOLACODE{attacks} is a list of the attack
data of the $14$ different attacks that the monitor can detect.
The nested specification \INNERNAME{flowAnalyzer} can use
retroactive dynamic parameterization or (the less efficient)
non-retroactive dynamic parameterization.

The second specification analyzes individual flows:

\noindent {\frontendsize\tt
\begin{tabularx}{\textwidth}{@{}r@{\hspace{.5em}}l}
\multicolumn{2}{X}{\color[HTML]{888888}\rule[2mm]{30mm}{.1pt}} \\
\linensize{\color[HTML]{4d798f}1} & \textbf{type }\textcolor[HTML]{008000}{\textbf{Entropy }}\textbf{=} Map.Map String (Set.Set String) \\
\linensize{\color[HTML]{4d798f}2} & \textbf{type }\textcolor[HTML]{008000}{\textbf{Histogram }}\textbf{=} Map.Map String Int \\
\linensize{\color[HTML]{4d798f}3} & \textcolor[HTML]{606060}{-{}- packets, bits, starttime, endtime} \\
\linensize{\color[HTML]{4d798f}4} & \textbf{type }\textcolor[HTML]{008000}{\textbf{AddressInfo }}\textbf{=} Map.Map String (Int, Int, Int, Int) \\[0.5em]
\linensize{\color[HTML]{4d798f}5} & \textbf{input} \textcolor[HTML]{008000}{\textbf{String}} \textcolor[HTML]{eb7600}{\textbf{fileid}} \\
\linensize{\color[HTML]{4d798f}6} & \textbf{input} \textcolor[HTML]{008000}{\textbf{Flow}} \textcolor[HTML]{eb7600}{\textbf{flow}} \\[0.5em]
\linensize{\color[HTML]{4d798f}7} & \textbf{define} \textcolor[HTML]{008000}{\textbf{Int}} \textcolor[HTML]{eb7600}{\textbf{flowCounter}} \textbf{=} flowCounter\textbf{[}-1\textbf{|}0\textbf{]} + 1 \\[0.5em]
\linensize{\color[HTML]{4d798f}8} & \textbf{define} \textcolor[HTML]{008000}{\textbf{Bool}} \textcolor[HTML]{eb7600}{\textbf{firstFlow}} \textbf{=} fileId\textbf{[now]} /= fileid \textbf{[}-1\textbf{|}""\textbf{]} \\
\linensize{\color[HTML]{4d798f}9} & \textbf{define} \textcolor[HTML]{008000}{\textbf{Bool}} \textcolor[HTML]{eb7600}{\textbf{lastFlow}} \textbf{=} fileid\textbf{[now]} /= fileId\textbf{[}1\textbf{|}""\textbf{]} \\[0.5em]
\linensize{\color[HTML]{4d798f}10} & \textbf{output} \textcolor[HTML]{008000}{\textbf{[String]}} \textcolor[HTML]{eb7600}{\textbf{attacked\_IPs}} \textbf{=} map detect attacks \\
\linensize{\color[HTML]{4d798f}11} & \phantom{..}\textcolor[HTML]{1c21b8}{where} detect atk = (attack\_detection atk)\textbf{[now]} \\[0.5em]
\linensize{\color[HTML]{4d798f}12} & \textbf{define} \textcolor[HTML]{008000}{\textbf{String}} \textcolor[HTML]{eb7600}{\textbf{attack\_detection}} \textbf{<}\textcolor[HTML]{008000}{\textbf{AttackData}} \textcolor[HTML]{96325d}{\textbf{atk}}\textbf{>} \textbf{=}  \\
\linensize{\color[HTML]{4d798f}13} & \phantom{..}\textcolor[HTML]{1c21b8}{if} (markerRate atk)\textbf{[now]} > threshold atk \textcolor[HTML]{1c21b8}{then} \\
\linensize{\color[HTML]{4d798f}14} & \phantom{..}\phantom{..}\textcolor[HTML]{1c21b8}{if} (ipEntropy atk)\textbf{[now]} > maxEntropy atk \textcolor[HTML]{1c21b8}{then} \\
\linensize{\color[HTML]{4d798f}15} & \phantom{..}\phantom{..}\phantom{..}(maxDestAddress atk)\textbf{[now]} \\
\linensize{\color[HTML]{4d798f}16} & \phantom{..}\phantom{..}\textcolor[HTML]{1c21b8}{else} "Over threshold but not entropy" \\
\linensize{\color[HTML]{4d798f}17} & \phantom{..}\textcolor[HTML]{1c21b8}{else} "No attack"\\
\multicolumn{2}{X}{\color[HTML]{888888}\rule[2mm]{30mm}{.1pt}}
\end{tabularx}}

\noindent {\frontendsize\tt
\begin{tabularx}{\textwidth}{@{}r@{\hspace{.5em}}l}
\multicolumn{2}{X}{\color[HTML]{888888}\rule[2mm]{30mm}{.1pt}} \\
\linensize{\color[HTML]{4d798f}18} & \textbf{define} \textcolor[HTML]{008000}{\textbf{Int}} \textcolor[HTML]{eb7600}{\textbf{markerRate}} \textbf{<}\textcolor[HTML]{008000}{\textbf{AttackData}} \textcolor[HTML]{96325d}{\textbf{atk}}\textbf{>} \textbf{=} \\
\linensize{\color[HTML]{4d798f}19} & \phantom{..}\textcolor[HTML]{1c21b8}{if} timeDur == 0 \textcolor[HTML]{1c21b8}{then} 0 \\
\linensize{\color[HTML]{4d798f}20} & \phantom{..}\textcolor[HTML]{1c21b8}{else} (getMarker addrData)\textbf{[now]} `div` timeDur \\
\linensize{\color[HTML]{4d798f}21} & \phantom{..}\textcolor[HTML]{1c21b8}{where} \\
\linensize{\color[HTML]{4d798f}22} & \phantom{..}\phantom{..}timeDur = getTE addrData - geTS addrData \\
\linensize{\color[HTML]{4d798f}23} & \phantom{..}\phantom{..}addrData = (addrInfo atk)\textbf{[now]} ! (maxDestAddress atk)\textbf{[now]} \\[0.5em]
\linensize{\color[HTML]{4d798f}24} & \textbf{define} \textcolor[HTML]{008000}{\textbf{String}} \textcolor[HTML]{eb7600}{\textbf{maxDestAddress}} \textbf{<}\textcolor[HTML]{008000}{\textbf{AttackData}} \textcolor[HTML]{96325d}{\textbf{atk}}\textbf{>} \textbf{=} \\
\linensize{\color[HTML]{4d798f}25} & \phantom{..}\textcolor[HTML]{1c21b8}{if} occurrencesCurrent > occurrencesPrev \\
\linensize{\color[HTML]{4d798f}26} & \phantom{..}\textcolor[HTML]{1c21b8}{then} currentAddr \textcolor[HTML]{1c21b8}{else} previousMaxAddr \\
\linensize{\color[HTML]{4d798f}27} & \phantom{..}\textcolor[HTML]{1c21b8}{where} \\
\linensize{\color[HTML]{4d798f}28} & \phantom{..}\phantom{..}currentAddr = destAddr\textbf{[now]} \\
\linensize{\color[HTML]{4d798f}29} & \phantom{..}\phantom{..}currentHist = (attackHist atk)\textbf{[now]} \\
\linensize{\color[HTML]{4d798f}30} & \phantom{..}\phantom{..}occurrencesCurrent = currentHist ! currentAddr \\
\linensize{\color[HTML]{4d798f}31} & \phantom{..}\phantom{..}occurrencesPrev = currentHist ! previousMaxAddr \\
\linensize{\color[HTML]{4d798f}32} & \phantom{..}\phantom{..}previousMaxAddr = (maxDestAddress atk)\textbf{[}-1\textbf{|}""\textbf{]} \\[0.5em]
\linensize{\color[HTML]{4d798f}33} & \textbf{define} \textcolor[HTML]{008000}{\textbf{AddrInfo}} \textcolor[HTML]{eb7600}{\textbf{addrInfo}} \textbf{<}\textcolor[HTML]{008000}{\textbf{AttackData}} \textcolor[HTML]{96325d}{\textbf{atk}}\textbf{>} \textbf{=} \\
\linensize{\color[HTML]{4d798f}34} & \phantom{..}insertWith updateValue destAddr\textbf{[now]} flow\textbf{[now]} prev \\
\linensize{\color[HTML]{4d798f}35} & \phantom{..}\textcolor[HTML]{1c21b8}{where} \\
\linensize{\color[HTML]{4d798f}36} & \phantom{..}\phantom{..}prev = \textcolor[HTML]{1c21b8}{if} firstFlow\textbf{[now]} \textcolor[HTML]{1c21b8}{then} empty \textcolor[HTML]{1c21b8}{else} (addrInfo atk) \textbf{[}-1\textbf{|}empty\textbf{]}\phantom{..}\phantom{..}  \\
\linensize{\color[HTML]{4d798f}37} & \phantom{..}\phantom{..}updateValue (p,b,ts,te) (p',b',ts',te') = (p+p',b+b',min ts ts',max te te') \\[0.5em]
\linensize{\color[HTML]{4d798f}38} & \textbf{define} \textcolor[HTML]{008000}{\textbf{Histogram}} \textcolor[HTML]{eb7600}{\textbf{attackHist}} \textbf{<}\textcolor[HTML]{008000}{\textbf{AttackData}} \textcolor[HTML]{96325d}{\textbf{atk}}\textbf{>} \textbf{=} \\
\linensize{\color[HTML]{4d798f}39} & \phantom{..}insertWith (+) destAddr\textbf{[now]} 1 hist \\
\linensize{\color[HTML]{4d798f}40} & \phantom{..}\textcolor[HTML]{1c21b8}{where} \\
\linensize{\color[HTML]{4d798f}41} & \phantom{..}\phantom{..}hist = \textcolor[HTML]{1c21b8}{if} firstFlow\textbf{[now]} \textcolor[HTML]{1c21b8}{then} empty \textcolor[HTML]{1c21b8}{else} (attackHist atk) \textbf{[}-1\textbf{|}empty\textbf{]}\phantom{..}\phantom{..} \\
\multicolumn{2}{X}{\color[HTML]{888888}\rule[2mm]{30mm}{.1pt}}
\end{tabularx}}

\noindent{} The specification defines many auxiliary types to carry
information about the destination addresses.
It also needs to keep the \STRNAME{flowCounter} to perform retroactive
dynamic parameterization.
As in the previous case, the stream \STRNAME{attacked\_IPs} maps the
parametric stream \STRNAME{attack\_detection} over the list of
attacks.
The stream \STRNAME{attack\_detection} checks that the marker (bits
per seconds or packets per second) of the attack and the IP entropy of
any address do not exceed the thresholds.
If the thresholds are exceeded, the IP address most accessed (which is
calculated in \STRNAME{maxDestAddress}) is considered to be under
attack.
The stream \STRNAME{markerRate} calculates the bits per seconds or
packets per second of an attack.
The stream \STRNAME{maxDestAddress} calculates the most accessed
address, comparing the accesses of the current destination address
(\LOLACODE{destAddr\NOW}) with the accesses of the previously most
accessed address (\LOLACODE{(maxDestAddress atk)\LOLAACC{-1}{""}}).
The binary operator \LOLACODE{(!)} retrieves an element from a map.
The stream \STRNAME{addrInfo} keeps a map of the packets, bits,
start time and endtime per destination address.
Similarly, the stream \STRNAME{attackHist} keeps a map of the number
of accesses per destination address.

The IP entropy used in line $14$ can be defined using brute force, or
retroactice dynamic parameterization.
We will first the see brute force version.

\noindent {\frontendsize\tt
\begin{tabularx}{\textwidth}{@{}r@{\hspace{.5em}}l}
\multicolumn{2}{X}{\color[HTML]{888888}\rule[2mm]{30mm}{.1pt}} \\
\linensize{\color[HTML]{4d798f}1} & \textbf{define} \textcolor[HTML]{008000}{\textbf{Int}} \textcolor[HTML]{eb7600}{\textbf{ipEntropy}} \textbf{<}\textcolor[HTML]{008000}{\textbf{AttackData}} \textcolor[HTML]{96325d}{\textbf{atk}}\textbf{>} \textbf{=} \\
\linensize{\color[HTML]{4d798f}2} & \phantom{..}size ((ipEntropyAllAddr atk)\textbf{[now]} ! (maxDestAddress atk)\textbf{[now]}) \\[0.5em]
\linensize{\color[HTML]{4d798f}3} & \textbf{define} \textcolor[HTML]{008000}{\textbf{Entropy}} \textcolor[HTML]{eb7600}{\textbf{ipEntropyAllAddr}} \textbf{<}\textcolor[HTML]{008000}{\textbf{AttackData}} \textcolor[HTML]{96325d}{\textbf{atk}}\textbf{>} \textbf{=} \\
\linensize{\color[HTML]{4d798f}4} & \phantom{..}insertWith union destAddr\textbf{[now]} (singleton srcAddr\textbf{[now]}) prevEntropy \\
\linensize{\color[HTML]{4d798f}5} & \phantom{..}\textcolor[HTML]{1c21b8}{where} \\
\linensize{\color[HTML]{4d798f}6} & \phantom{..}\phantom{..}prevEntropy = \textcolor[HTML]{1c21b8}{if} firstFlow\textbf{[now]} \textcolor[HTML]{1c21b8}{then} empty \\
\linensize{\color[HTML]{4d798f}7} & \phantom{..}\phantom{..}\phantom{..}\phantom{..}\phantom{..}\phantom{..}\phantom{..}\phantom{..}\phantom{..}\textcolor[HTML]{1c21b8}{else} (ipEntropyAllAddr atk) \textbf{[}-1\textbf{|}empty\textbf{]}\phantom{..}\phantom{..} \\
\multicolumn{2}{X}{\color[HTML]{888888}\rule[2mm]{30mm}{.1pt}}
\end{tabularx}}

\noindent{} In this specification we calculate the ip entropy of every
address at all times, and we simply return the size of the set of
different origin IP addresses of the most accesseed IP.
The entropy calculation that uses retroactive parameterization is
the following.

\noindent {\frontendsize\tt
\begin{tabularx}{\textwidth}{@{}r@{\hspace{.5em}}l}
\multicolumn{2}{X}{\color[HTML]{888888}\rule[2mm]{30mm}{.1pt}} \\
\linensize{\color[HTML]{4d798f}1} & \textbf{define} \textcolor[HTML]{008000}{\textbf{Int}} \textcolor[HTML]{eb7600}{\textbf{ipEntropy}} \textbf{<}\textcolor[HTML]{008000}{\textbf{AttackData}} \textcolor[HTML]{96325d}{\textbf{atk}}\textbf{>} \textbf{=} \\
\linensize{\color[HTML]{4d798f}2} & \phantom{..}maybe 0 size mset \\
\linensize{\color[HTML]{4d798f}3} & \phantom{..}\textcolor[HTML]{1c21b8}{where} \\
\linensize{\color[HTML]{4d798f}4} & \phantom{..}\phantom{..}mset = setSrcForDestAddr atk \\
\linensize{\color[HTML]{4d798f}5} & \phantom{..}\phantom{..}\phantom{..}\phantom{..}\phantom{..} `mover` maybeAddress attData \\
\linensize{\color[HTML]{4d798f}6} & \phantom{..}\phantom{..}\phantom{..}\phantom{..}\phantom{..} `withInit` initer atk (maxDestAddress atk)\textbf{[now]} flowCounter\textbf{[now]} \\[0.5em]
\linensize{\color[HTML]{4d798f}7} & \textbf{define} \textcolor[HTML]{008000}{\textbf{(Set String)}} \textcolor[HTML]{eb7600}{\textbf{setSrcForDestAddr}} \textbf{<}\textcolor[HTML]{008000}{\textbf{AttackData}} \textcolor[HTML]{96325d}{\textbf{atk}}\textbf{>} \textbf{<}\textcolor[HTML]{008000}{\textbf{String}} \textcolor[HTML]{96325d}{\textbf{dst}}\textbf{>} \textbf{=} \\
\linensize{\color[HTML]{4d798f}8} & \phantom{..}insert srcAddr\textbf{[now]} prevSet \\
\linensize{\color[HTML]{4d798f}9} & \phantom{..}\textcolor[HTML]{1c21b8}{where} \\
\linensize{\color[HTML]{4d798f}10} & \phantom{..}\phantom{..}prevSet = \textcolor[HTML]{1c21b8}{if} firstFlow\textbf{[now]} \textcolor[HTML]{1c21b8}{then} empty \\
\linensize{\color[HTML]{4d798f}11} & \phantom{..}\phantom{..}\phantom{..}\phantom{..}\phantom{..}\phantom{..}\phantom{..}\textcolor[HTML]{1c21b8}{else} (setSrcForDestAddr atk) \textbf{[}-1\textbf{|}empty\textbf{]}\phantom{..}\phantom{..}  \\[0.5em]
\linensize{\color[HTML]{4d798f}12} & \textbf{define} \textcolor[HTML]{008000}{\textbf{(Maybe String)}} \textcolor[HTML]{eb7600}{\textbf{maybeAddress}} \textbf{<}\textcolor[HTML]{008000}{\textbf{AttackData}} \textcolor[HTML]{96325d}{\textbf{atk}}\textbf{>} \textbf{=} \\
\linensize{\color[HTML]{4d798f}13} & \phantom{..}\textcolor[HTML]{1c21b8}{if} (maxOverThreshold atk)\textbf{[now]} \\
\linensize{\color[HTML]{4d798f}14} & \phantom{..}\textcolor[HTML]{1c21b8}{then} Just (maxDestAddress atk)\textbf{[now]} \\
\linensize{\color[HTML]{4d798f}15} & \phantom{..}\textcolor[HTML]{1c21b8}{else} Nothing\\
\multicolumn{2}{X}{\color[HTML]{888888}\rule[2mm]{30mm}{.1pt}}
\end{tabularx}}

\noindent{} In this case, we define a parametric stream
\STRNAME{setSrcForDestAddr} that calculates the set of different
origin IPs of a destination address.
We define an auxiliary stream \STRNAME{maybeAddress} that contains the
most accessed address, if it exceeds the threshold.
The definition of \STRNAME{ipEntropy} will instantiate dynamically the
stream \STRNAME{setSrcForDestAddr} with the most accessed address once
it exceeds the threshold, with an initializer specific to the
suspected attack and address.

\end{document}